\documentclass[preprint]{revtex4}
\usepackage{graphicx}
\usepackage{subfigure}
\usepackage{multirow}
\usepackage{hyperref}
\usepackage{amsmath,bm,amssymb,amsfonts}
\usepackage{color}
\newcommand{\dalm}{\kern1pt\vbox{\hrule height 0.9pt\hbox{\vrule width 0.9pt
			\hskip 2.5pt\vbox{\vskip 5.5pt}\hskip 3pt\vrule width 0.3pt}\hrule height 0.3pt}
	\kern1pt}
\begin{document}
	\title{{\bf
			Spontaneous Scalarization in Proto-neutron Stars}}
	
	\author{{\bf Fahimeh Rahimi $^{1,2}$ } and {\bf Zeinab Rezaei $^{1,2}$} \footnote{Corresponding author. E-mail:
				zrezaei@shirazu.ac.ir}}
	\affiliation{ $^{1}$Department of Physics, School of Science, Shiraz
University, Shiraz 71454, Iran.\\
		$^{2}$Biruni Observatory, School of Science, Shiraz
University, Shiraz 71454, Iran.}
	
	%%%%%%%%%%%%%%%%%%%%%%%%%%%%%%%%%%%%%%%%%%%%%%%%%%%%%%%%%%%%%%%%%%%%%%%%%
	
	\begin{abstract}

Proto-neutron stars are born when a highly evolved and massive star collapses under gravity. In this paper, we investigate the spontaneous scalarization in proto-neutron stars. Based on the scalar tensor theory of gravity as well as the physical conditions in proto-neutron star, we examine the
structure of proto-neutron star. To describe the fluid in proto-neutron star, we utilize $SU(2)$ chiral sigma model and
the finite temperature extension of the Brueckner-Bethe-Goldstone quantum many-body theory
in the Brueckner-Hartree-Fock approximation. Here, we apply the equation of state of proto-neutron stars considering different cases i.e. hot pure neutron matter and hot $\beta$-stable neutron star matter without neutrino trapping as well as with neutrino trapping. The effects of temperature and entropy of proto-neutron stars on the star structure are also studied. Our results confirm that the spontaneous scalarization is affected by different physical conditions in proto-neutron stars.

	\end{abstract}
	
	\maketitle
	%%%%%%%%%%%%%%%%%%%%%%%%%%%%%%%%%%%%%%%%%%%%%%%%%%%%%%%

	\section{INTRODUCTION}
Proto-neutron stars (PNSs), the hot objects in the first stages of the neutron star (NS) evolution, are the most important laboratories for relativistic astrophysics.
A PNS undergoes several stages and cooling down by neutrino emission leading to a stable NS.
Different methods are employed to understand how nuclear matter properties depend on the temperature.
Equation of state of nuclear matter at finite temperature describing supernova of NSs has been presented using the relativistic mean field theory  \cite{shen1998relativistic,shen2011second}.
In Brueckner-Bethe-Goldstone theoretical approach, the minimum as well as the maximum mass of PNS has been explored \cite{burgio2010maximum}.
Self-consistent mean-field theory has been applied to study the latent heat of the liquid-gas phase transition in symmetric nuclear matter \cite{carbone2011latent}.
By applying microscopic two- and three-body nuclear potentials derived from chiral effective field theory, the effect of density and temperature on the nuclear symmetry free energy has been considered \cite{wellenhofer2015thermodynamics}.
Brueckner-Hartree-Fock approach at finite temperature including three-body forces has been employed to investigate
the properties of hot $\beta$-stable nuclear matter \cite{lu2019hot}.
In the effective chiral model, the finite temperature properties of nuclear matter have been studied \cite{sen2020nuclear}.

Neutrinos in PNSs can affect the structural properties of these stars.
The number of trapped neutrinos in PNS has important role in determining the composition of PNSs \cite{prakash1997composition}.
Finite temperature Brueckner-Bethe-Goldstone theoretical approach confirms that finite temperature and neutrino trapping reduce the value of the PNS maximum mass
 \cite{nicotra2006protoneutron}.
Properties of neutrino-heated winds from PNSs have been studied \cite{neut1,neut16}.
Neutrino radiation has significant effects on the momentum, heat, and lepton transports in PNSs \cite{neut15}.
Minimum and maximum mass and the radius of PNSs have been studied in hadronic chiral SU(3) model considering the trapped neutrinos \cite{neut14}.
Charged-current neutrino interactions in hot and dense matter affect the luminosities and
spectra of the emitted neutrinos \cite{neut13}.
Relativistic mean field calculations verify that asymmetric neutrino absorption influences the spin deceleration of PNSs \cite{neut12}.
Neutrino signal in core-collapse supernovae can be applied to determine the time evolution of PNS structure \cite{neut4,neut7,neut10,neut11}.
Two-point correlation function for pairs of neutrinos can employ to estimate the size of the PNS \cite{neut8}.
Simulation with neutrino transport and proper motions of PNS shows that the asymmetric neutrino emissions
have significant effects on the acceleration of PNS \cite{neut6}.
Charged current neutrino processes affect the formation and evolution of neutrino spectra during the deleptonization of PNSs \cite{neut9}.
Applying a quasi-static approach based on parameterized entropy and electron fraction profiles evolving with neutrino cooling processes,
the mass and radius of PNSs have been calculated \cite{neut5}.
Neutrinos emitted anisotropically from the PNS can generate gravitational waves during the cooling phase \cite{neut2}.
Utilizing the full set of charged-current neutrino-nucleon reactions, the Kelvin-Helmholtz cooling phase of PNSs has been simulated \cite{neut3}.

One of the most interesting consequences of scalar tensor theory is that NSs can undergo what is known as scalarization.
Damour and Esposito-Farese investigated stellar structure in tensor-scalar theories \cite{damour1993nonperturbative}. In their study, they discovered a phenomenon similar to spontaneous magnetization in ferromagnetism in low temperatures, which has been called spontaneous scalarization.
Spontaneous scalarization depends on the star matter equation of state \cite{whinnett2004new}.
Spontaneous scalarization takes place within theories in which the effective linear coupling $\beta$ between the scalar and matter
fields is sufficiently negative \cite{mendes2016highly}. However, a similar instability to tachyonic-like can also occur for the values of the coupling constant $\beta>0$ in sufficiently compact neutron stars. The final state of this instability depends on the coupling function where gravitational collapse or spontaneous scalarization can take place \cite{mendes2016highly}.
Weakening of gravity by spontaneous scalarization as well as scalar force alters the internal structure of stars \cite{STT3}.
In the scalar tensor Gauss Bonnet theories, the curvature of spacetime leads to the star scalarization \cite{doneva2018neutron}.
Theory parameter space for spontaneous scalarization has been calculated applying the linearized scalar field equation \cite{STT6}.
Calculations in the tensor-multi-scalar theories of gravity result in the scalarized branches characterized by the number of the scalar field nodes \cite{STT2}.
Scalarization in charged stars reduces the Coulomb force and results in stars with smaller masses and radii \cite{STT1}.
Spontaneous scalarization alters the magnetic deformation of NSs and the emission of quadrupolar gravitational waves \cite{STT5}.
NS mass and radius measurements can constrain the spontaneous scalarization parameters in scalar tensor theories \cite{STT4}.

According to the above discussions, it is obvious that the spontaneous scalarization can take place in PNSs. Here, we study the effects of PNS equation of state on this scalarization. In section \ref{s2}, we present the equations of state describing the PNSs.
Section \ref{s3} summarizes the scalar tensor theory of gravity which leads to the spontaneous scalarization of PNSs.
In Section \ref{s4}, the mass and radius of PNSs are discussed. The scalar field in PNSs is described in Section \ref{s5}.
Section \ref{s6} belongs to the scalar charge of PNSs. Summary and conclusions are given in section \ref{s7}.

	\section{Proto-neutron Star Equations of State}\label{s2}
In this paper, we apply two formalisms i.e. $SU(2)$ chiral sigma model \cite{sahu2004hot} and
the finite temperature extension of the Brueckner-Bethe-Goldstone quantum many-body theory \cite{pnseos2}
to describe the PNS equation of state. In the chiral sigma model, the dense matter equation of state is analyzed within the variation of different parameters which are in agreement with the results obtained in the heavy-ion collision experiments. In this model, equation of state is calculated applying the conserved total stress tensor related to the Lagrangian and the mean-field equation of motion
for the Fermion field and a mean-field approximation for the meson fields \cite{sahu2004hot}.
\begin{figure*}
\vspace*{1cm}       % Give the correct figure height in cm
\includegraphics[width=0.5\textwidth]{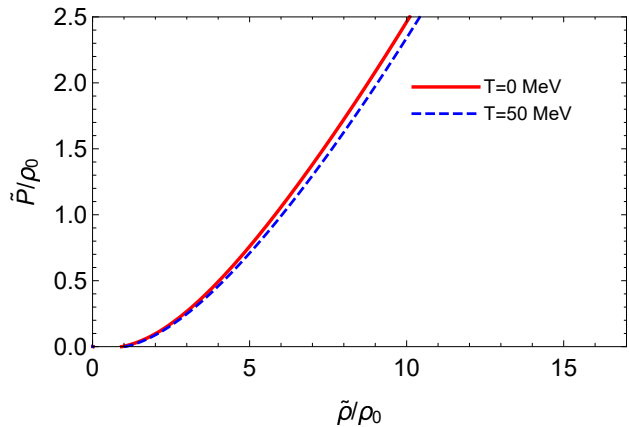}
\caption{Pure neutron matter equation of state in the chiral sigma model at two temperatures \cite{sahu2004hot}, $\rho_0=1.66\times10^{14}g/cm^3$.}
\label{p-rhot0}
\end{figure*}
In Fig. \ref{p-rhot0}, we have presented the equation of state of pure neutron matter in the chiral sigma model at two temperatures with $\rho_0=1.66\times10^{14}g/cm^3$.

\begin{figure}[h]
		\subfigure{}{\includegraphics[scale=0.85]{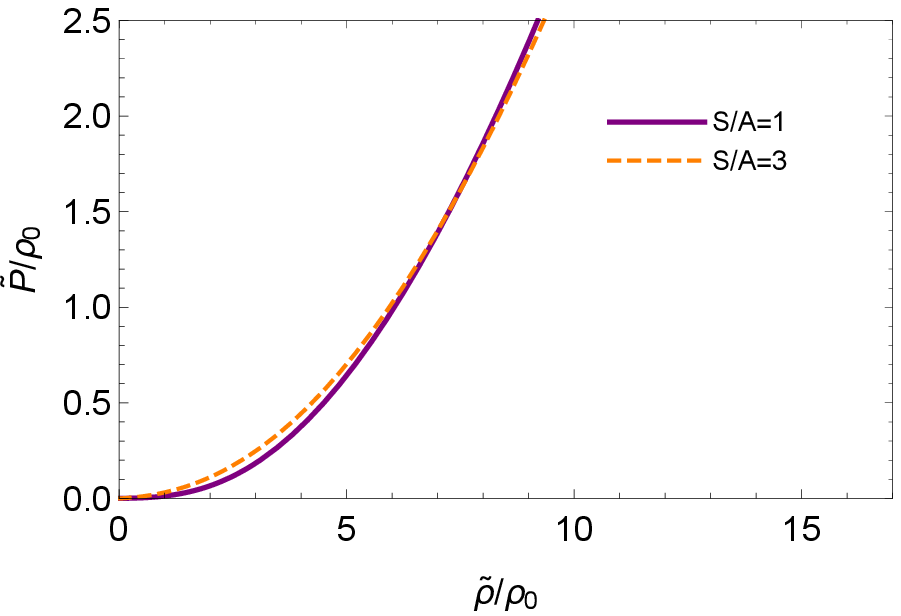}
		\subfigure{}{\includegraphics[scale=0.85]{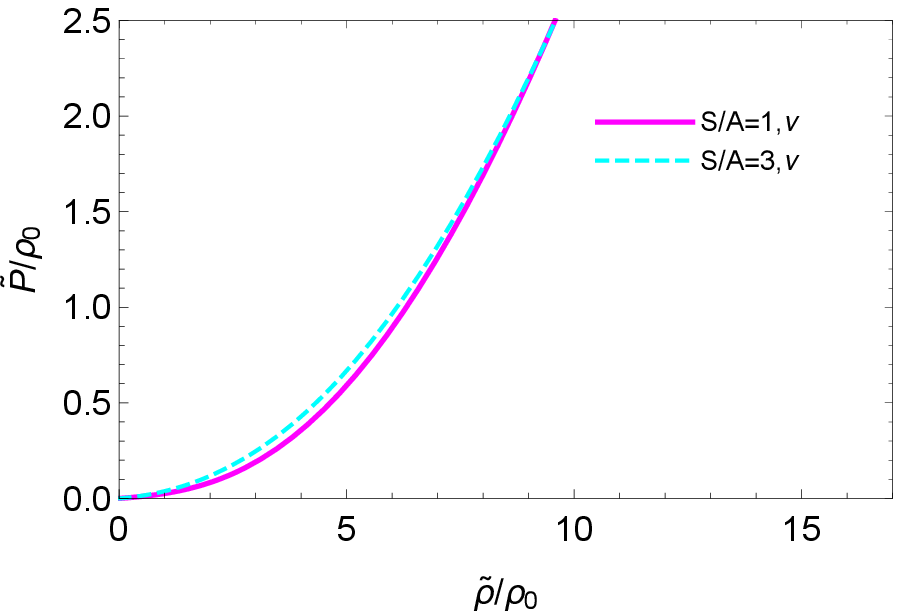}}}	
		\caption{Left: Equation of state of $\beta$-stable neutron star matter at different values of the entropy per baryon $S/A$ in the case of neutrino-less \cite{pnseos2},          Right: Same as Left but for the case of neutrino trapped matter (denoted by $\nu$) \cite{pnseos2}, $\rho_0=1.66\times10^{14}g/cm^3$.}
		\label{p-rhoS}
	\end{figure}

Recently, for modeling the relativistic astrophysical environments such as PNSs, a microscopic equation of state for hot $\beta$-stable nuclear matter has been presented \cite{pnseos2}. This equation of state has been calculated employing the finite temperature extension of the Brueckner-Bethe-Goldstone quantum many-body theory in the Brueckner-Hartree-Fock approximation. In this approach, the contributions of massive leptons, photons, and possibly neutrinos ($\nu$) in three different species have been added to the hadronic entropy, energy density, and pressure.
Due to high temperatures of PNSs, a large number of neutrinos are produced in the star. Neutrinos initially form a trapped neutrino gas
and diffuse out over the diffusion timescales \cite{pnseos2}. Fig. \ref{p-rhoS} shows the equation of state of isoentropic $\beta$-stable neutron star matter at different values of the entropy per baryon $S/A$ considering the cases of neutrino-less and neutrino trapped matter \cite{pnseos2}.

	\section{Scalar Tensor Theory of Gravity}\label{s3}
	We consider the class of scalar-tensor theories in Einstein frame which can be defined by the following action \cite{mendes2016highly},
	\begin{equation}
		S[g_{\mu \nu},\phi,\Psi_m] = \frac{1}{16 \pi} \int d^4x \sqrt{-g}(R-2 \nabla_\mu\phi \nabla^\mu \phi)+S_m[\Psi_m,a(\phi)^2 g_{\mu \nu}],
	\end{equation}
	where $g=det(g_{\mu \nu})$, $R$ is Ricci scalar, $\phi$ denotes scalar field, and $\Psi_m$ is matter field. In addition, $a(\phi)$ is coupling function that relates $g_{\mu \nu}$ in Einstein frame and $\tilde{g} _{\mu \nu}$ in Jordan frame by conformal transformation as $\tilde{g} _{\mu \nu}=a(\phi)g_{\mu \nu}$.
In this paper, we consider two coupling function models as $a(\phi) = e^{\frac{1}{2}\beta (\phi-\phi_0) ^2}$ (Model 1) and $a(\phi) =[ \cosh \left(\sqrt{3} \beta (\phi-\phi_0)\right)]^\frac{1}{3\beta}$ (Model 2) \cite{damour1993nonperturbative,mendes2016highly}, in which $\beta$ is the coupling constant and  $\phi_0=0$.
Considering the spherically symmetric static spacetime line element, $ds^2 = - N(r)^2 dt^2 + A(r)^2 dr^2 + r^2 (d\theta^2 + \sin^2\theta d\phi^2)$, where $A(r) = [1-2 m(r)/r ]^{-1/2}$ and $m(r)$ is the mass profile, field equations result in the following differential equations \cite{mendes2016highly},
	\begin{align}
		&\frac{d m}{dr} = 4\pi r^2 a^4 \tilde{\epsilon} + \frac{r}{2} (r-2m) \Big(\frac{d\phi}{dr}\Big)^2 \label{eq:dm},\\
		&\frac{d \ln N}{dr} = \frac{4\pi r^2 a^4 \tilde{p}}{r - 2m} +\frac{r}{2} \Big(\frac{d\phi}{dr}\Big)^2 + \frac{m}{r(r-2m)} \label{eq:dn}, \\
		&\frac{d^2\phi}{dr^2} = \frac{4\pi r a^4}{r-2m} \! \left[ \alpha (\tilde{\epsilon} - 3\tilde{p}) + r (\tilde{\epsilon} - \tilde{p}) \frac{d\phi}{dr} \right ]\! -\frac{2(r-m)}{r(r-2m)} \frac{d\phi}{dr} \label{eq:dphi}, \\
		&\frac{d\tilde{p}}{dr} = -(\tilde{\epsilon} + \tilde{p}) \left[  \frac{4\pi r^2 a^4 \tilde{p}}{r-2m} \! + \! \frac{r}{2} \Big(\frac{d\phi}{dr}\Big)^2 \!\! + \! \frac{m}{r(r-2m)} \! + \! \alpha \frac{d\phi}{dr} \right], \label{eq:dp}\\
		&\frac{dm_b}{dr}=\frac{4\pi r^2a^3\tilde{\rho}}{\sqrt{1-\frac{2m}{r}}}\label{eq:dm_b},
	\end{align}
with energy density $\tilde{ \epsilon}$, pressure $\tilde{p}$, and baryonic mass $m_b$. These equations are numerically solved by applying the boundary conditions,
\begin{align}
		&m(0) =m_b(0)= 0, \quad \lim_{r\to\infty}N(r) = 1,\quad \phi(0)=\phi_c, \quad \lim_{r\to\infty}\phi(r) = 0, \nonumber \\
		&\frac{d\phi}{dr}(0) = 0, \qquad \tilde{p}(0) = p_c, \qquad \tilde{p}(R_s) = 0, \label{eq:bc}
	\end{align}
	where $R_s$ is the radius of the star and $c$ denotes the center of star. We use a fourth-order Runge-Kutta algorithm based on appropriate boundary conditions at $r=0$ and a guess $\phi(0)=\phi_c$ at the center and iterating on $\phi_c$ to get the following condition \cite{damour1993nonperturbative,mendes2016highly}, \begin{equation} \label{eq:constraint on central of scalar field}
		\phi_s  + \frac{2 \psi_s}{\sqrt{\dot{\nu}_s^2+4\psi_s^2}} \textrm{arctanh} \left[ \frac{\sqrt{\dot{\nu}_s^2 +4\psi_s^2}}{\dot{\nu}_s +2/R_s} \right] = 0.
	\end{equation}
	The subscript s indicates quantities on the surface of the star. Besides, $\psi_s = (d\phi/dr)_s$ and $\dot{\nu}_s = 2(d\ln N/dr)|_s = R_s \psi_s^2 + 2 m_s/[R_s(R_s-2m_s)]$. Moreover, the ADM mass, $M_{ADM}$, and scalar charge, $\omega$, are given by \cite{damour1993nonperturbative,mendes2016highly},
	\begin{align}
		M_{ADM} &= \frac{R_s^2 \dot{\nu}_s}{2} \left( 1-\frac{2m_s}{R_s} \right)^\frac{1}{2}
		\exp \left[ \frac{-\dot{\nu}_s}{\sqrt{\dot{\nu}_s^2+4\psi_s^2}} \textrm{arctanh} \left( \frac{\sqrt{\dot{\nu}_s^2+4\psi_s^2}}{\dot{\nu}_s +2/R_s} \right) \right], \\
		\omega & = - 2 M_{ADM} \psi_s/\dot{\nu}_s.
	\end{align}
Here, we apply the above formalism to describe the structure of PNSs.

	\section{Proto-neutron star Mass}\label{s4}

\begin{figure}[h]
		\subfigure{}{\includegraphics[scale=0.85]{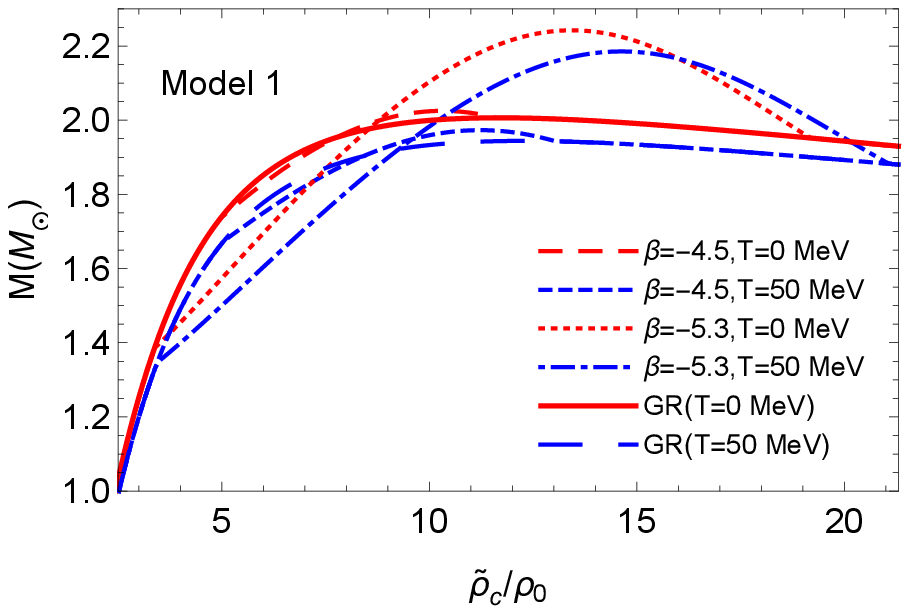}
				}
		\subfigure{}{\includegraphics[scale=0.85]{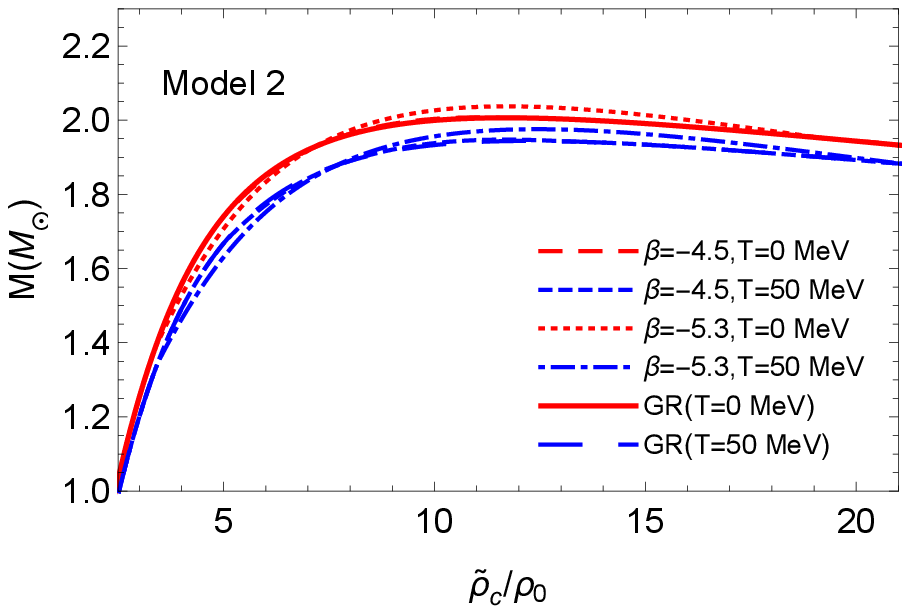}
				}	
		\caption{PNS mass, $M,$ versus the central density, $\rho_c,$ for cold NS and
			PNS at finite temperature with different values of the coupling constant, $\beta$, in two models for the coupling function. The values in general relativity (GR) are also given.}
		\label{Mro}
	\end{figure}

Fig. \ref{Mro} presents the PNS mass at finite temperature and the cold NS mass in two models for the coupling function and for different values of the coupling constant $\beta$.
The mass of PNS at finite temperature is lower than the one of cold NS.
For both cold NSs and PNSs at finite temperature, the deviation from general relativity (GR) is higher for the lower values of $\beta$.
The spontaneous scalarization is distinguishable in cold NSs and PNSs. Model 1 leads to larger deviations from GR.

	\begin{figure}[h]
		\subfigure{}{\includegraphics[scale=0.85]{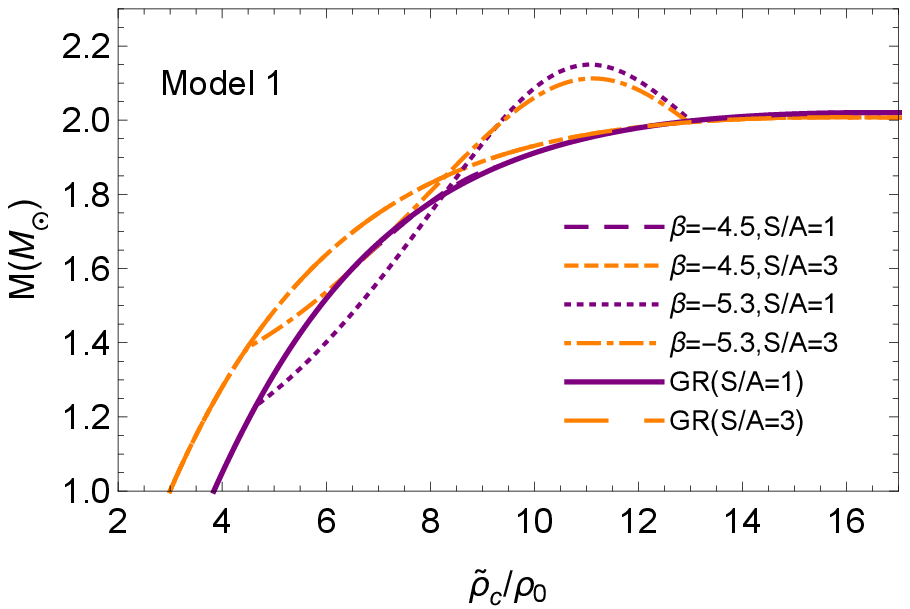}
				}
		\subfigure{}{\includegraphics[scale=0.85]{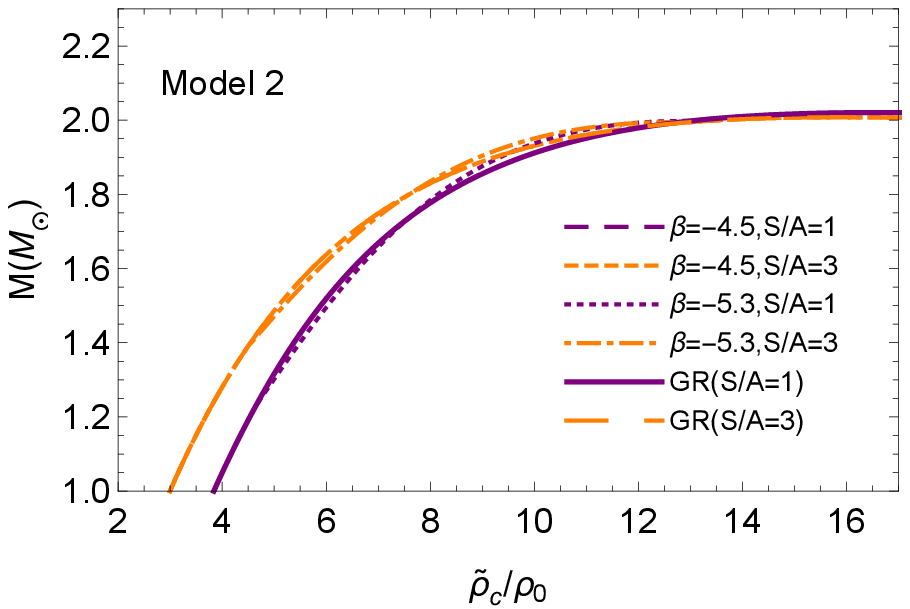}
			
			\subfigure{}{\includegraphics[scale=0.85]{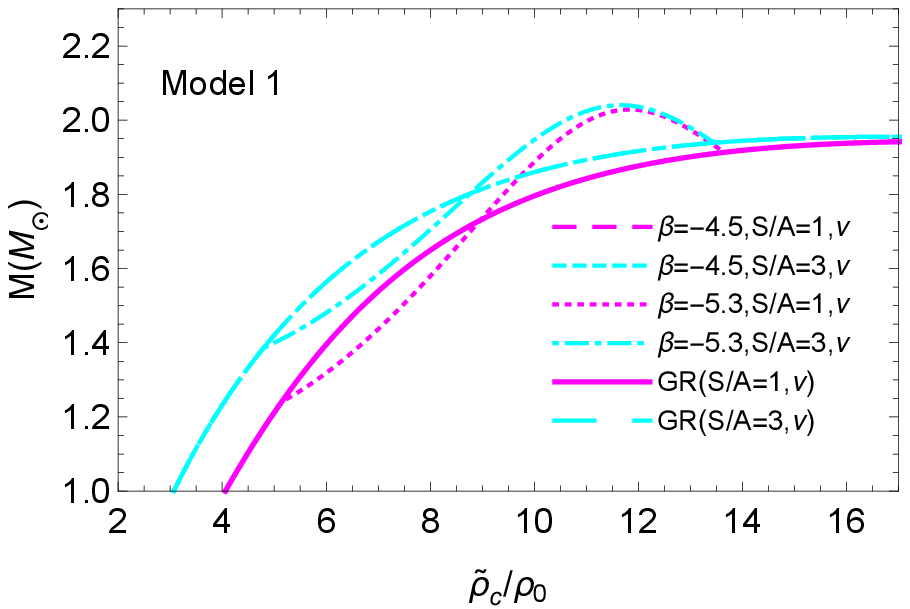}
				}
			\subfigure{}{\includegraphics[scale=0.85]{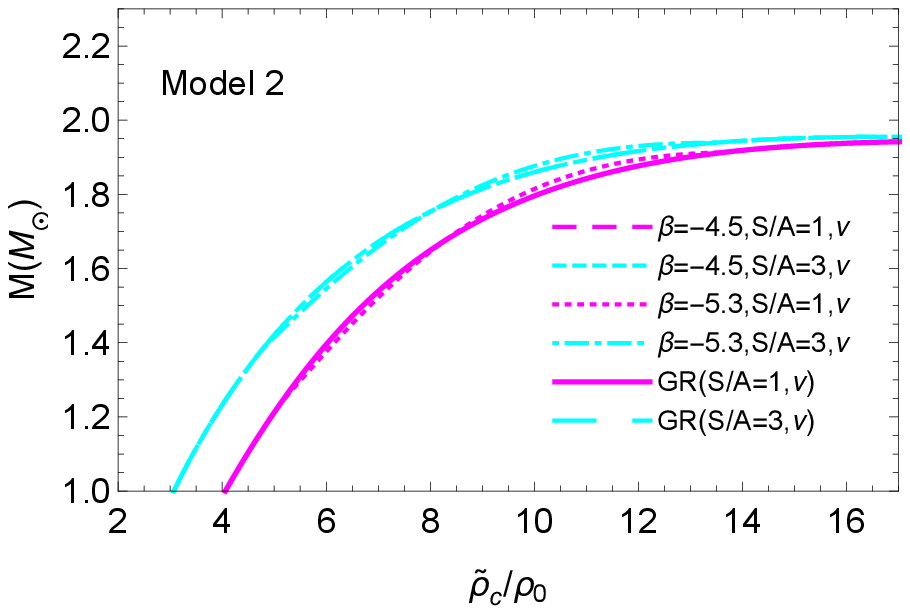}
				}	}	
		\caption{Same as Fig. \ref{Mro} but for PNSs with hot isoentropic $\beta$-stable neutron star matter without neutrino trapping and with neutrino trapping (denoted by $\nu$) for different values of entropy per baryon $S/A$.}
		\label{Mro2}
	\end{figure}

In Fig. \ref{Mro2}, we have shown the PNS mass with hot isoentropic $\beta$-stable neutron star matter in the cases without neutrino trapping as well as with neutrino trapping. Our results verify that for these PNSs with $\beta=-4.5$, the behavior of mass almost coincidences with GR. In most stars, higher values of the entropy result in more massive stars. This is while for some scalarized PNSs in the case without neutrino trapping, the mass decreases with increasing the entropy. In both cases without and with neutrino trapping, the spontaneous scalarization is slightly more significant for PNSs with lower values of entropy. For these PNSs,
the results of the scalar tensor gravity with Model 1 are more different from GR compared to Model 2.
Table \ref{table2:maximum mass} gives the maximum mass of PNSs for different cases which we have presented.

	\begin{table}[h!]
		\begin{center}
			\begin{tabular}{c@{\hspace{6mm}}c@{\hspace{6mm}}c@{\hspace{6mm}}c@{\hspace{6mm}}c@{\hspace{6mm}}c@{\hspace{6mm}}c@{\hspace{6mm}}c@{\hspace{6mm}}}
				\hline
				\hline
				& & Maximum Mass ($M_{\bigodot}$)& \\	
				\hline
				Model & $\beta$&T=0&T=50 MeV&S/A=1&S/A=3&S/A=1, $\nu$&S/A=3, $\nu$\\	
				\hline
				{1}&$-4.5$ &2.02 &1.97 &2.21 &2.01 &1.94 &1.96  \\
				&$-5.3$ &2.23 &2.19 &2.21 &2.11 &2.03 &2.04  \\
				\hline\hline
				{2}&$-4.5$ &2.01 &1.95 &2.02 &2.01 &1.94 &1.96  \\
				&$-5.3$ &2.04 &1.98 &2.02 &2.01 &1.94 &1.96  \\
				\hline\hline			
				GR & &2.01 & 1.94& 2.02&2.01&1.94&1.96\\			
				\hline\hline
			\end{tabular}
			
		\end{center}
		\caption{Maximum mass of cold NS and PNSs at finite temperature and with hot isoentropic $\beta$-stable neutron star matter without neutrino trapping and with neutrino trapping for different values of entropy per baryon $S/A$.}
		\label{table2:maximum mass}
	\end{table}

\begin{figure}[h]
		\subfigure{}{\includegraphics[scale=0.85]{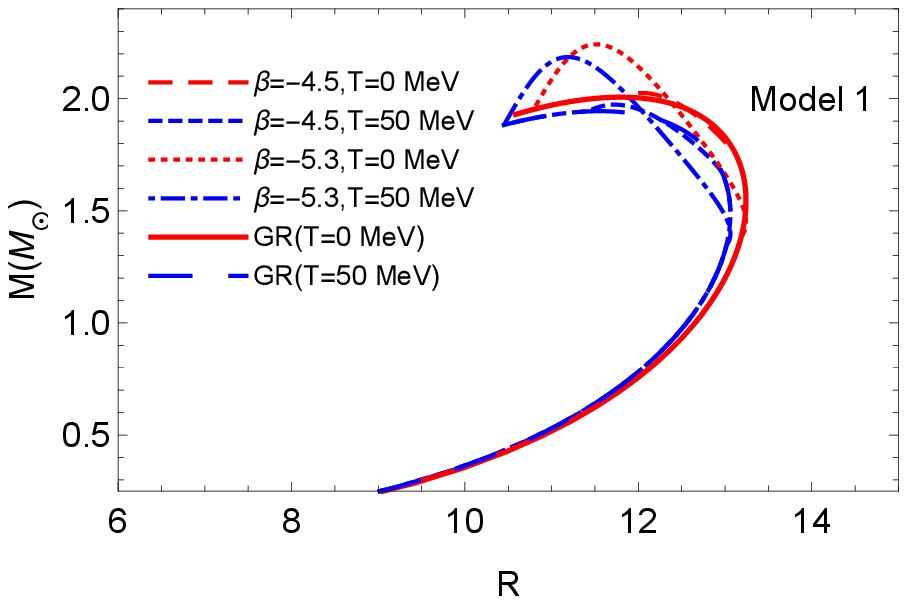}
			\label{mass-rho-1}}
		\subfigure{}{\includegraphics[scale=0.85]{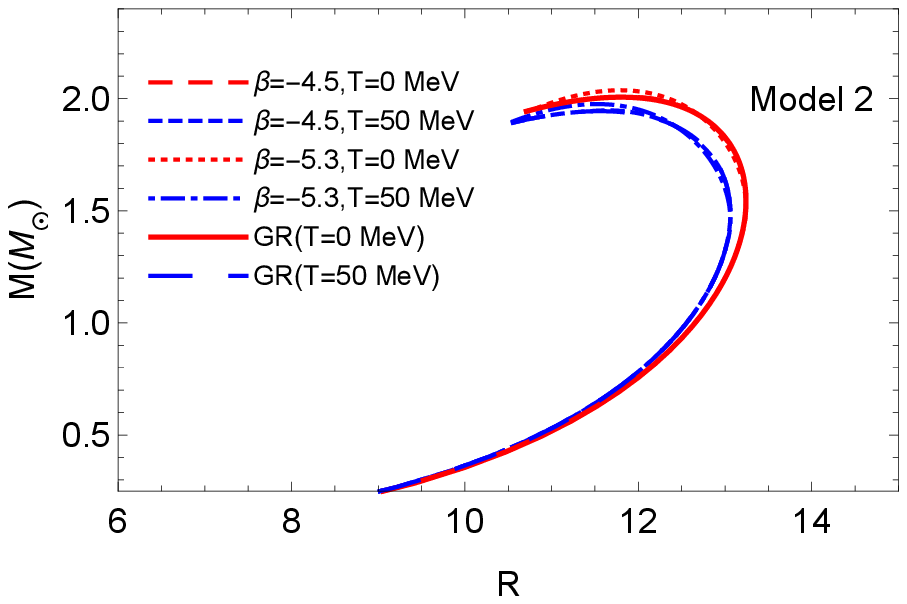}
			\label{mass-rho-2}}
		\subfigure{}{\includegraphics[scale=0.85]{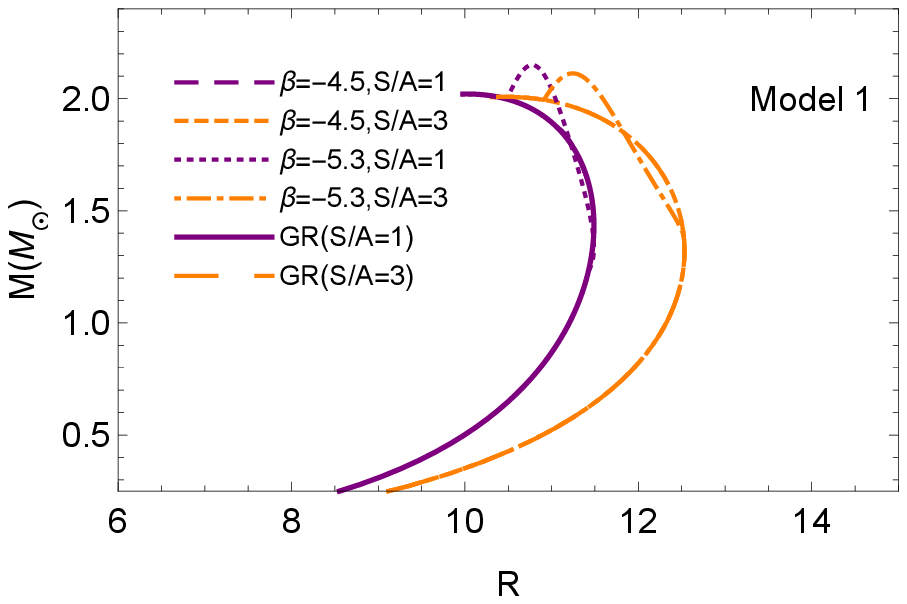}
			\label{phic-rho-1}}
		\subfigure{}{\includegraphics[scale=0.85]{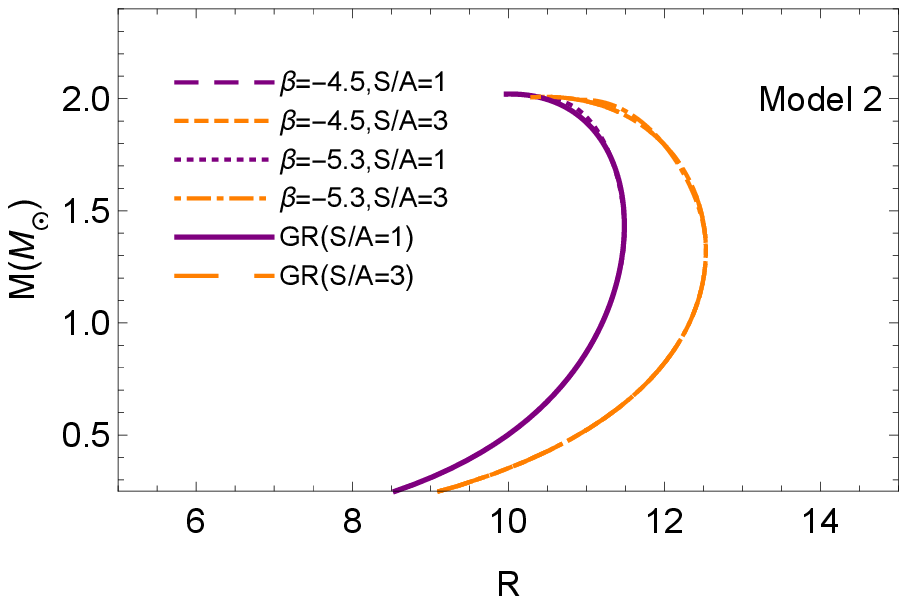}
			\label{phic-rho-2}}
		\subfigure{}{\includegraphics[scale=0.85]{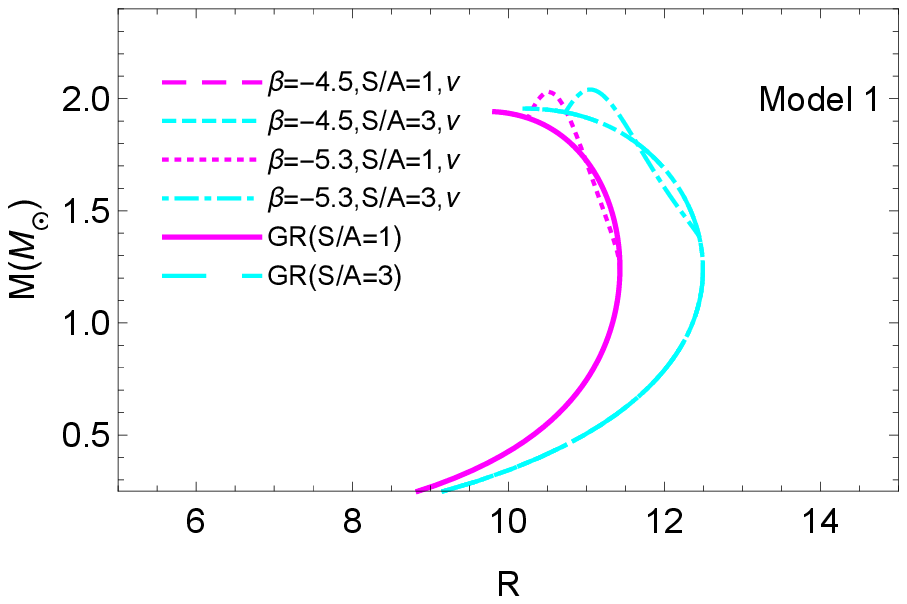}
			\label{phic-rho-1}}
		\subfigure{}{\includegraphics[scale=0.85]{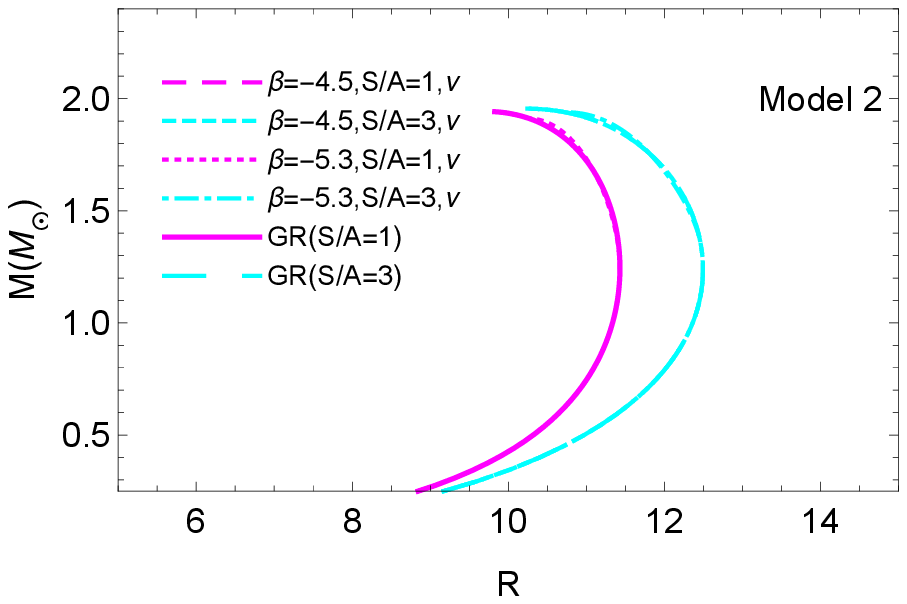}
			\label{phic-rho-2}}
		\caption{Mass versus the radius for cold NSs and PNSs with different equations of state and different values of the coupling constant in two models for the coupling function.}
		\label{3-mass-r1}
	\end{figure}

Fig. \ref{3-mass-r1} shows the mass-radius relation (MR relation) for cold NSs and PNSs. For both cold NS and PNS with $\beta=-5.3$ in Model 1, the deviation of MR relation from the GR one is obvious showing the spontaneous scalarization in these stars. Our results confirm that the PNSs at finite temperature are smaller in size compared to cold NSs. However, considering PNSs with hot isoentropic $\beta$-stable neutron star matter, the stars with higher values of the entropy are larger in size. PNSs at finite temperature experience more significant spontaneous scalarization in comparison with the ones containing hot isoentropic $\beta$-stable neutron star matter. The MR relations of scalarized PNSs are affected by the equations of state. MR relations in Model 2 are almost the same in the scalar tensor gravity and GR.

	\section{Scalar Field in Proto-neutron Star}\label{s5}

	\begin{figure}[h]
		\subfigure{}{\includegraphics[scale=0.85]{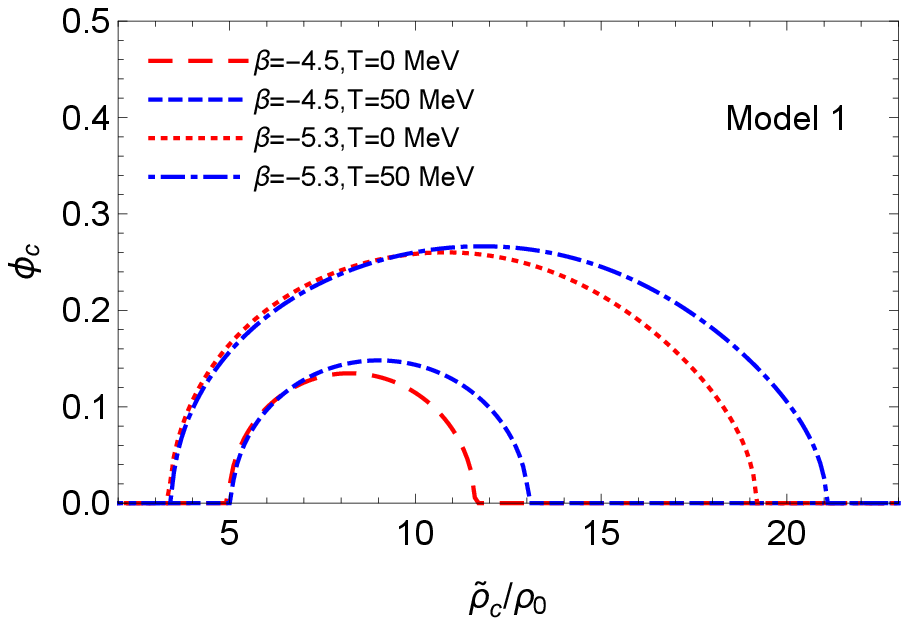}
			\label{phic-rho-11}}
		\subfigure{}{\includegraphics[scale=0.85]{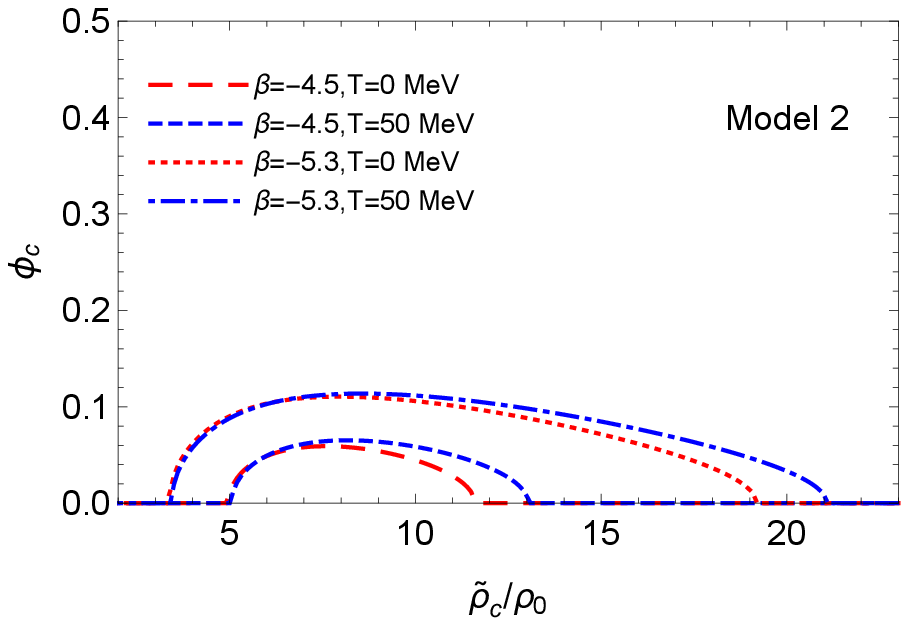}
			\label{phic-rho-22}}	
		\subfigure{}{\includegraphics[scale=0.85]{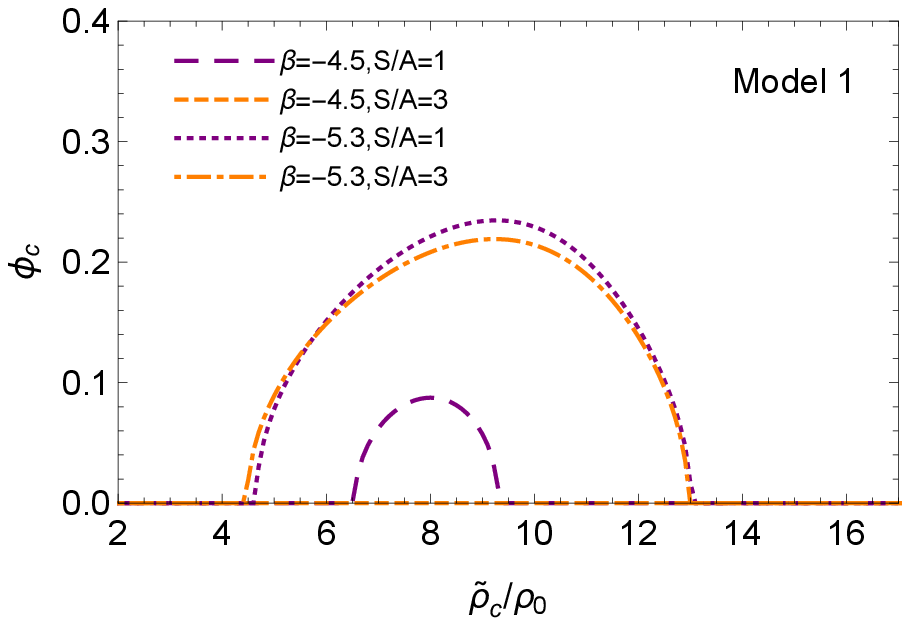}
			\label{phic-rho-3}}
		\subfigure{}{\includegraphics[scale=0.85]{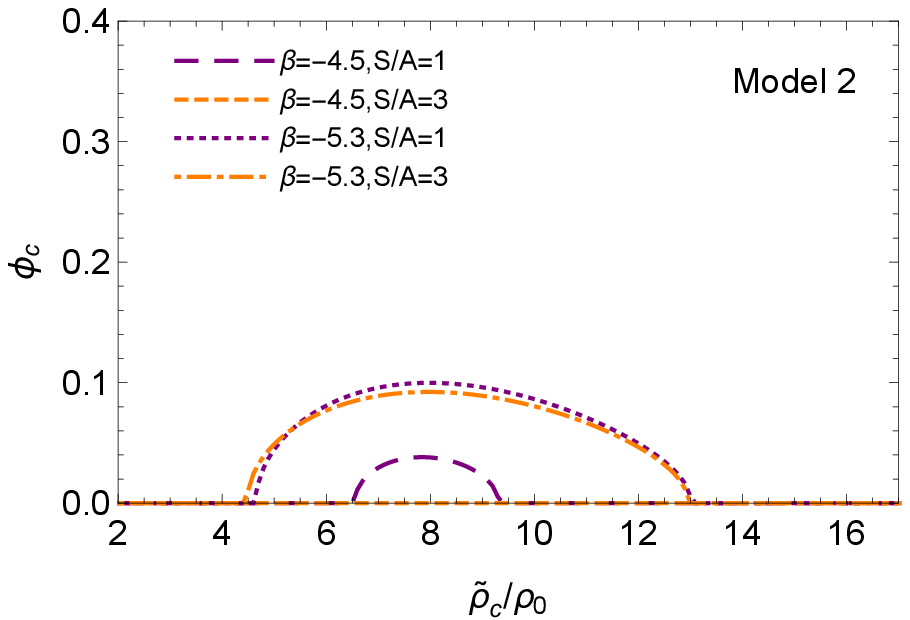}
			\label{phic-rho-4}}
		\subfigure{}{\includegraphics[scale=0.85]{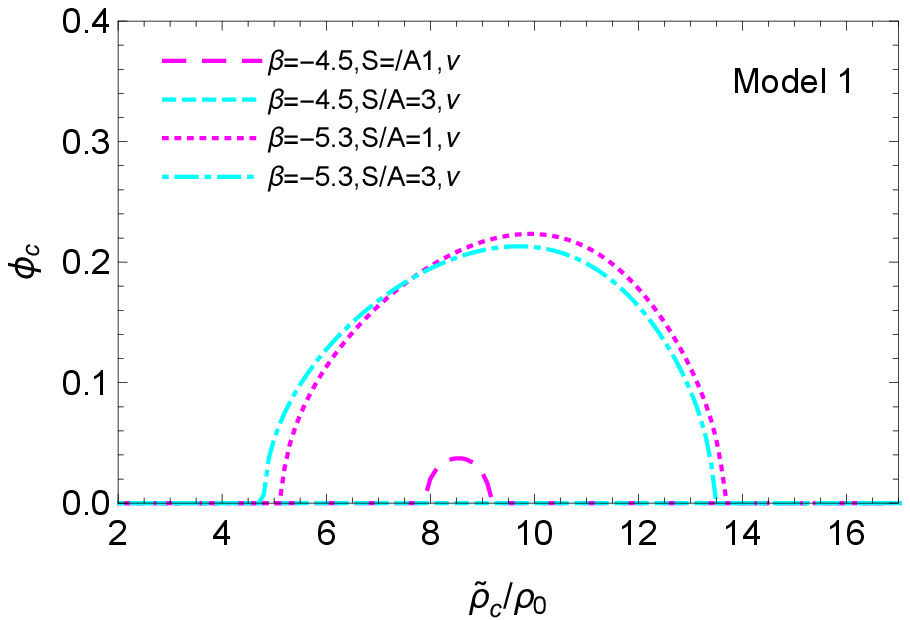}
			\label{phic-rho-5}}
		\subfigure{}{\includegraphics[scale=0.85]{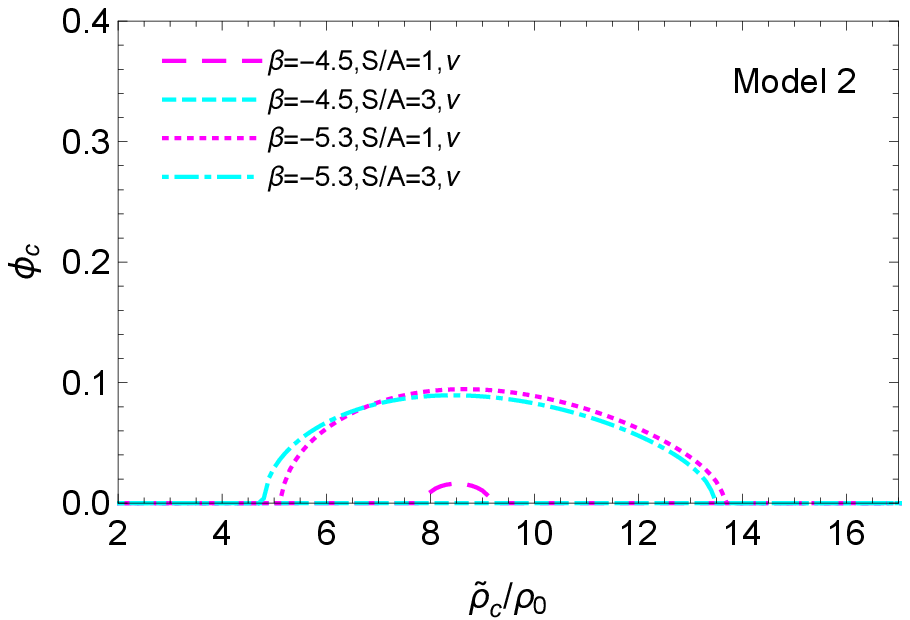}
			\label{phic-rho-6}}
		\caption{Central scalar field, $\phi_c$, versus the central density, $\rho_c$, for cold NSs and for PNSs
at finite temperature and with hot isoentropic $\beta$-stable neutron star matter without neutrino trapping and with neutrino trapping for different values of entropy per baryon $S/A$ and the coupling constant, $\beta$, in two models for the coupling function. Note that in GR we have $\phi_c=0$.}
		\label{3-phi-r11}
	\end{figure}

In Fig. \ref{3-phi-r11}, the central scalar field, $\phi_c$, in different stars is plotted as a function of the central density, $\rho_c$.
In all stars, $\phi_c$ increases as the coupling constant decreases. In the stars with hot isoentropic $\beta$-stable neutron star matter and $S/A=3$, the spontaneous scalarization does not take place for $\beta=-4.5$ and thus $\phi_c=0$ at each density.
Our calculations indicate that in order to appearance of spontaneous scalarization in these stars, the coupling constant should decrease to $\beta=-4.53$ and $\beta=-4.57$ in the cases of without and with neutrino trapping. In cold NSs and PNSs at finite temperature, the critical densities at which the spontaneous scalarization takes place (first critical density) are almost the same. This is while considering the stars with hot isoentropic $\beta$-stable neutron star matter, this critical density decreases as the entropy grows.
 Our results indicate that neutrino trapping leads to increase in the first critical density. PNSs at finite temperature can experience higher values of $\phi_c$ and the central density corresponding to maximum value of $\phi_c$ is larger in PNSs compared to cold NSs. In the case of stars with hot isoentropic $\beta$-stable neutron star matter, $\phi_c$ becomes larger at lower entropies.
 For these stars, the central density related to maximum central scalar field is higher for lower entropies. In hot isoentropic $\beta$-stable neutron star matter, neutrino trapping suppresses $\phi_c$ and the spontaneous scalarization.
Furthermore, the value of the central density corresponding to maximum value of $\phi_c$ is higher when neutrino trapping is considered.
At a special value of central density, the spontaneous scalarization terminates which we call that density second critical density. Considering the case with $\beta=-4.5$, the second critical density of PNSs at finite temperature is higher than that of cold NSs. For sufficiently low values of the coupling constant, e.g. $\beta=-5.3$, both cold NSs and PNSs at finite temperature are scalarazied even with the high values of the central density and the second critical density is greater than $19 \rho_0$. In PNSs with hot isoentropic $\beta$-stable neutron star matter without neutrino trapping, the second critical density does not depend on the entropy. In PNSs with neutrino trapping, the second critical density is larger compared to the case without neutrino trapping. Besides, the second critical density decreases as the entropy grows when the neutrino trapping is taken into account.

	\begin{figure}[h]
		\subfigure{}{\includegraphics[scale=0.85]{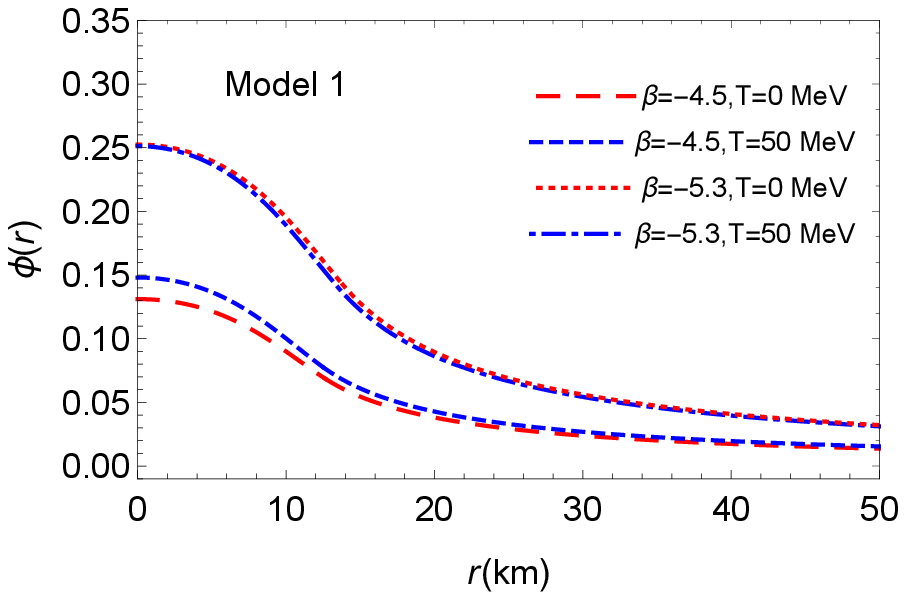}
			\label{phic-rho-1}}
		\subfigure{}{\includegraphics[scale=0.85]{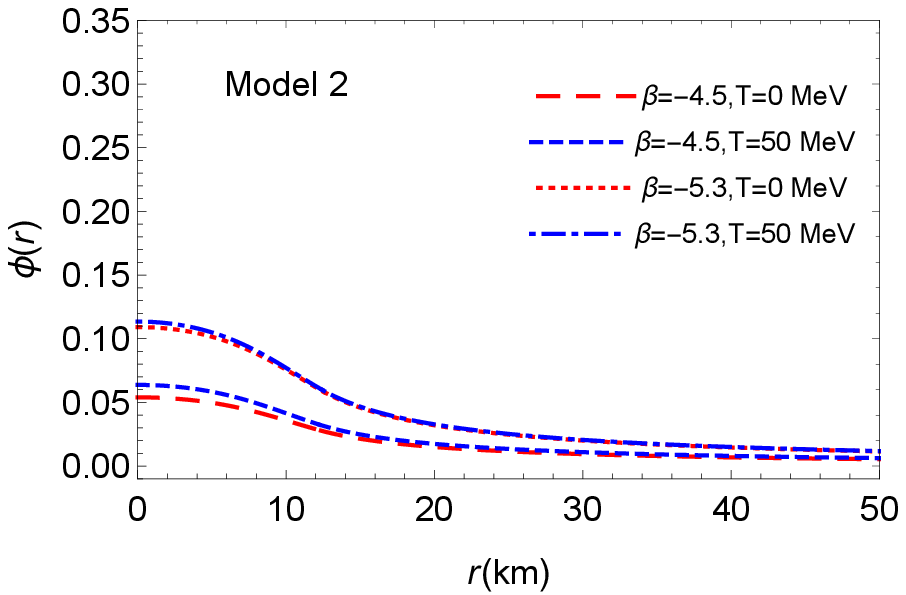}
			\label{phic-rho-2}}
		\subfigure{}{\includegraphics[scale=0.85]{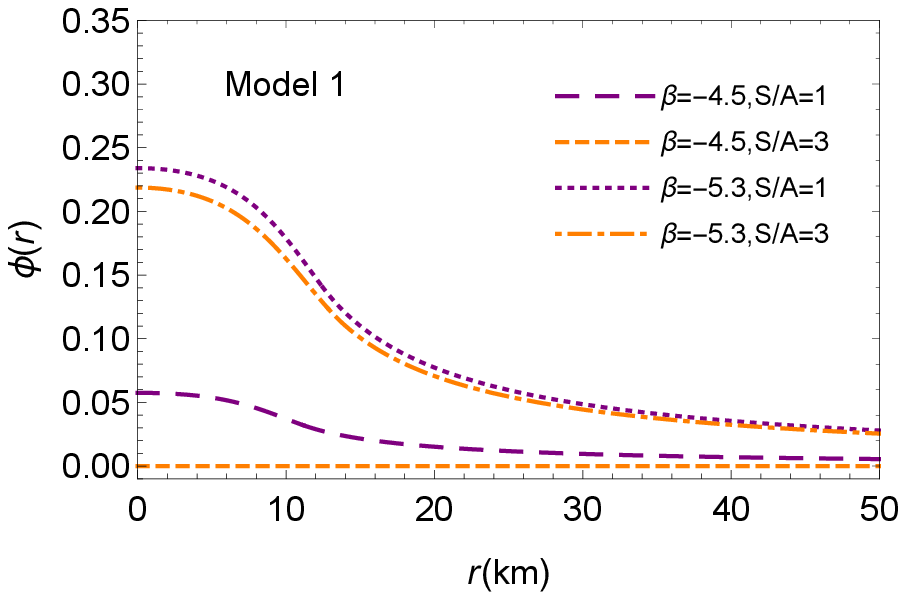}
			\label{phic-rho-1}}
		\subfigure{}{\includegraphics[scale=0.85]{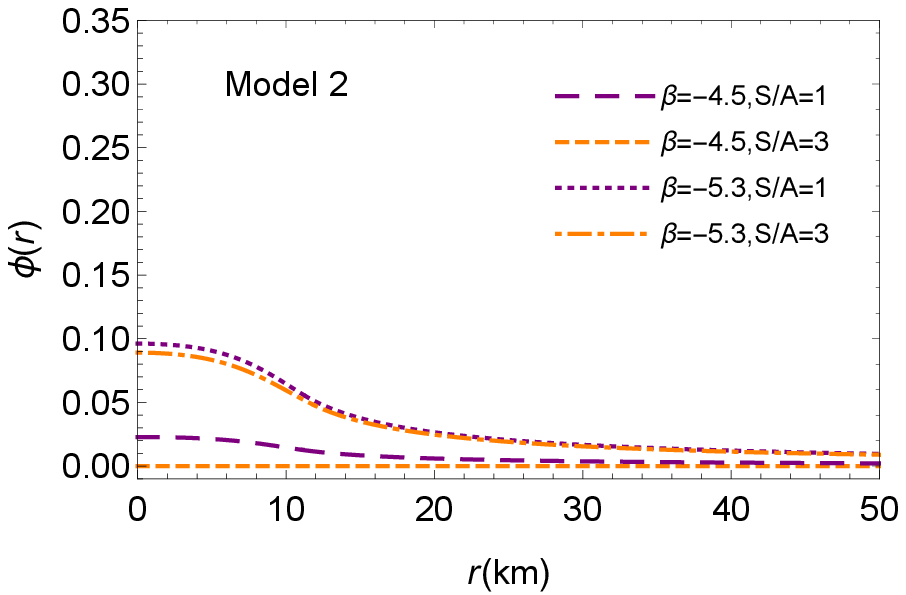}
			\label{phic-rho-2}}
		\subfigure{}{\includegraphics[scale=0.85]{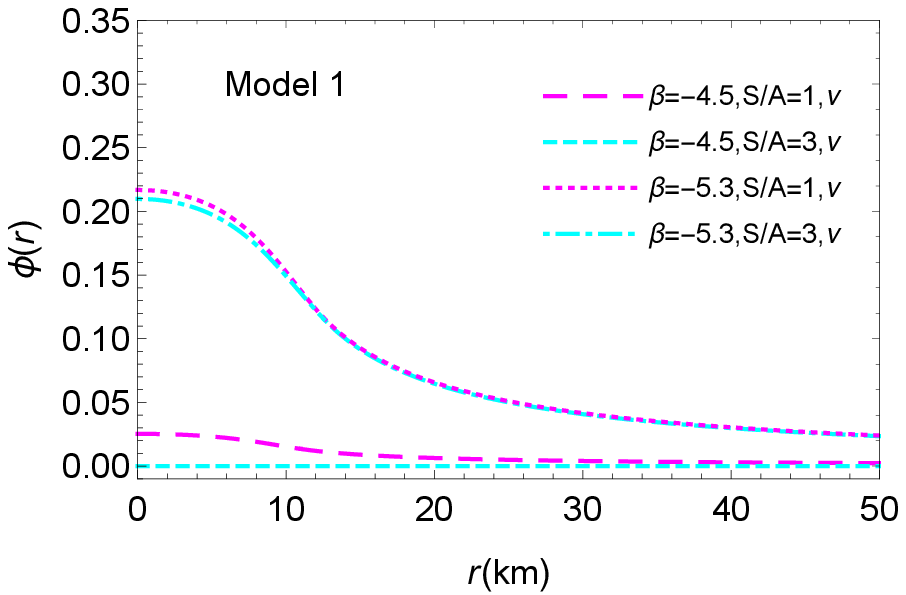}
			\label{phic-rho-1}}
		\subfigure{}{\includegraphics[scale=0.85]{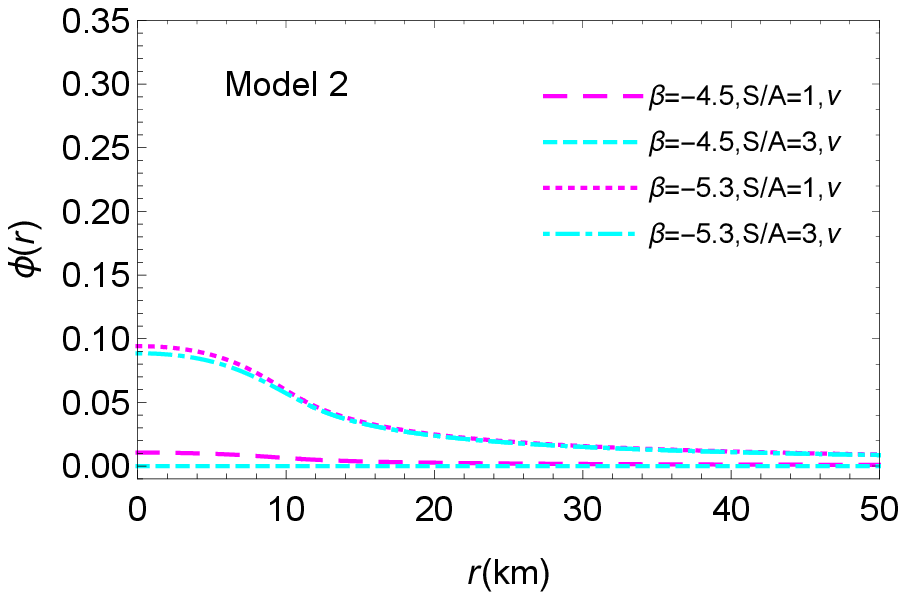}
			\label{phic-rho-2}}
		\caption{ Scalar field profiles in cold NSs and also in PNSs
at finite temperature and with hot isoentropic $\beta$-stable neutron star matter without neutrino trapping and with neutrino trapping for different values of entropy per baryon $S/A$ and the coupling constant, $\beta$, in two models for the coupling function. We have considered the value of the central density is $\rho_c=9\rho_0$.}
		\label{3-phi-r1}
	\end{figure}

Fig. \ref{3-phi-r1} shows the profile of scalar field in cold NSs and PNSs. In PNSs at finite temperature with lower values of $\beta$, the profile of scalar field behaves similar to the one in cold NSs, while with high values of $\beta$, the profile of PNSs is different from the one in cold NSs, specially at the center of the stars. In PNSs with hot isoentropic $\beta$-stable neutron star matter, the value of the scalar field profile reduces as the entropy increases. This reduction is more significant at the center of the stars. The effects of entropy on the scalar field profile are more important when $\beta$ is higher. In the case of PNSs with neutrino trapping, the profile of scalar field is less affected by the entropy than the one without neutrino trapping.
Out of the star, with lower values of $\beta$, the profiles of the scalar field are more similar compared to the ones with higher values of $\beta$. The effects of the coupling constant on the profile of the scalar field outside the star is more significant in Model 1.

\begin{figure*}
\vspace*{1cm}       % Give the correct figure height in cm
\includegraphics[width=0.5\textwidth]{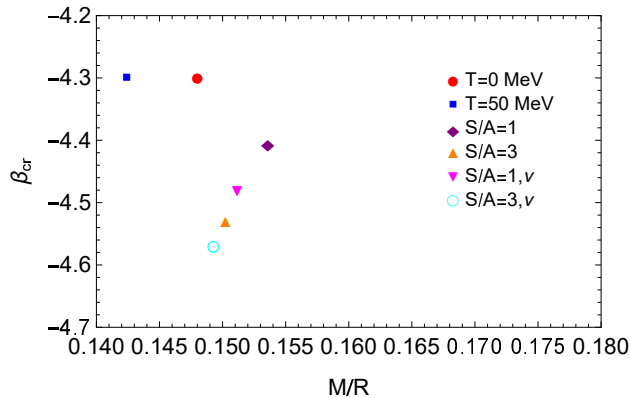}
\caption{Critical value of the coupling constant $\beta_{cr}$ as a function of the compactness $M/R$ in the cases of different equations of state considered in this work.}
\label{criticalbeta}
\end{figure*}

In Fig. \ref{criticalbeta}, we have presented the critical value of the coupling constant $\beta_{cr}$, the value of the coupling constant at which the scalarization appears. With the equation of state describing PNSs at finite temperature, $\beta_{cr}$ has the highest value with the lowest value of the corresponding compactness $M/R$. In the case of $S/A=1$ without neutrino trapping, the critical value of the coupling constant corresponds to highest value of the compactness $M/R$. Besides, considering $S/A=3$ with neutrino trapping, both $\beta_{cr}$ and the compactness have lower values.

	\section{Proto-neutron Star Scalar Charge}\label{s6}

	\begin{figure}[h]
		\subfigure{}{\includegraphics[scale=0.85]{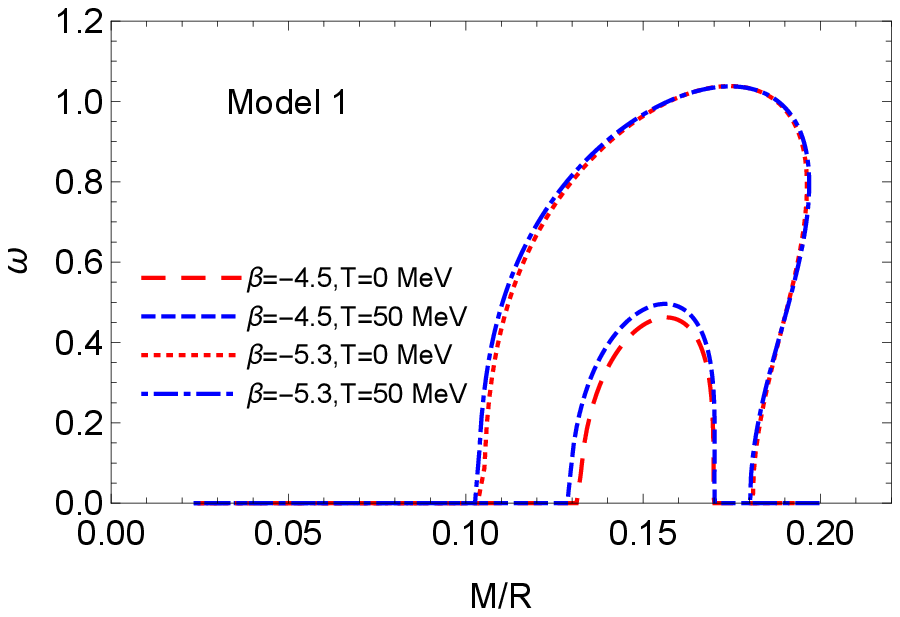}
			\label{phic-rho-1}}
		\subfigure{}{\includegraphics[scale=0.85]{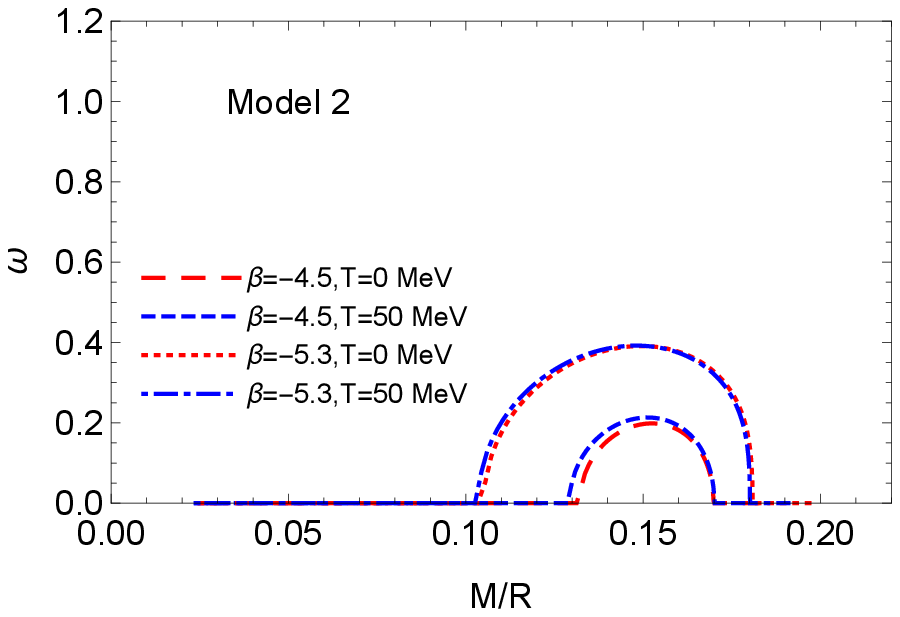}
			\label{phic-rho-2}}
		\subfigure{}{\includegraphics[scale=0.85]{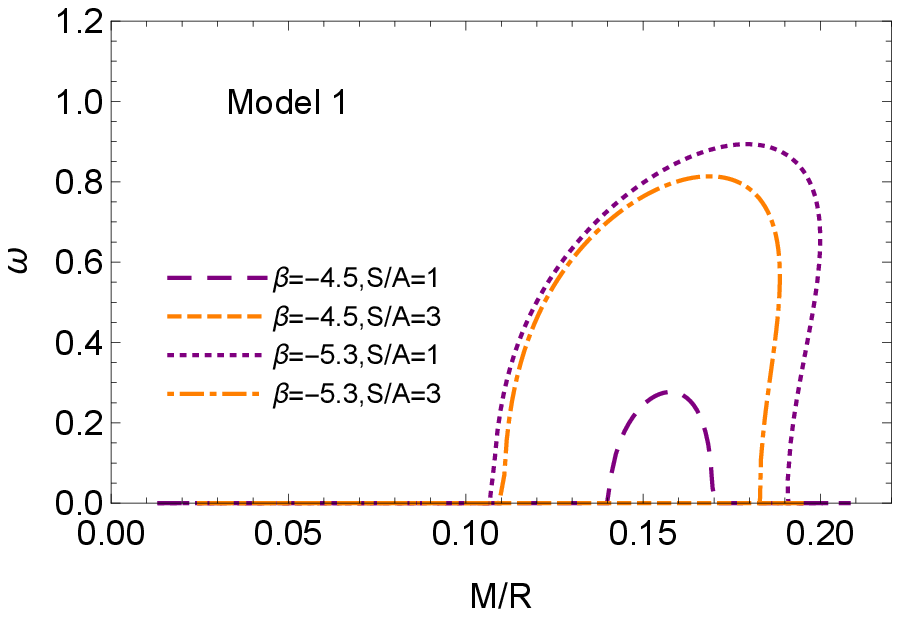}
			\label{phic-rho-1}}
		\subfigure{}{\includegraphics[scale=0.85]{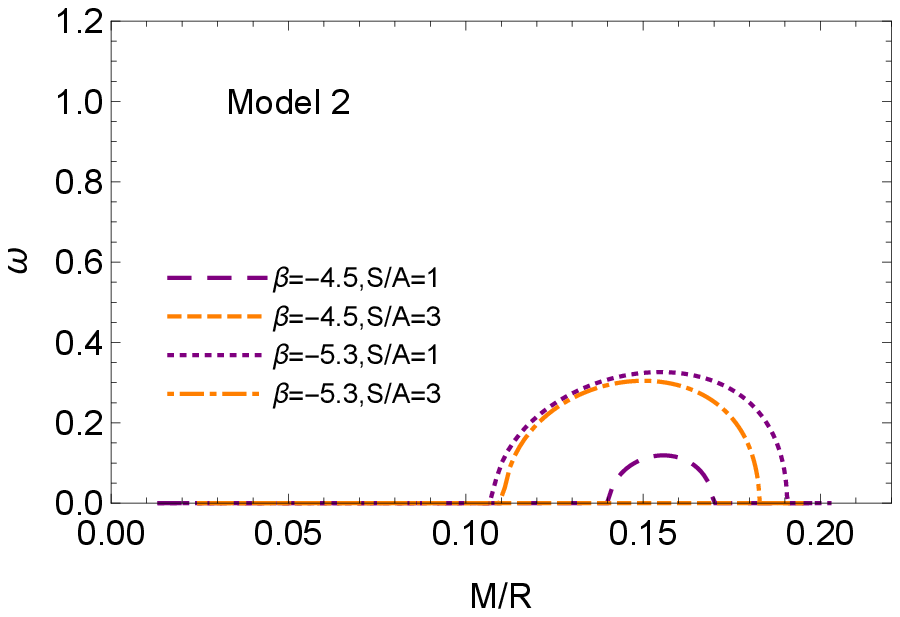}
			\label{phic-rho-2}}
		\subfigure{}{\includegraphics[scale=0.85]{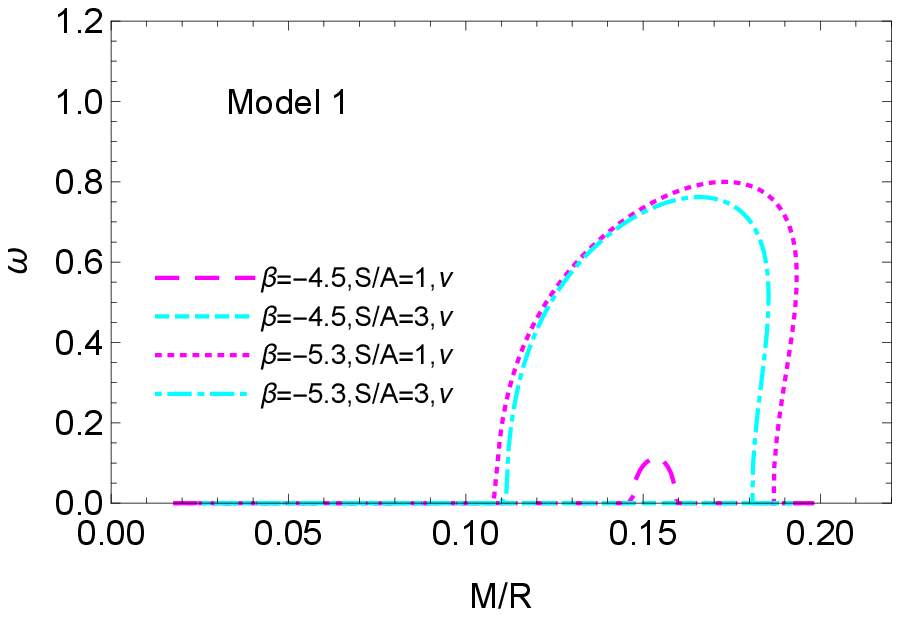}
			\label{phic-rho-1}}
		\subfigure{}{\includegraphics[scale=0.85]{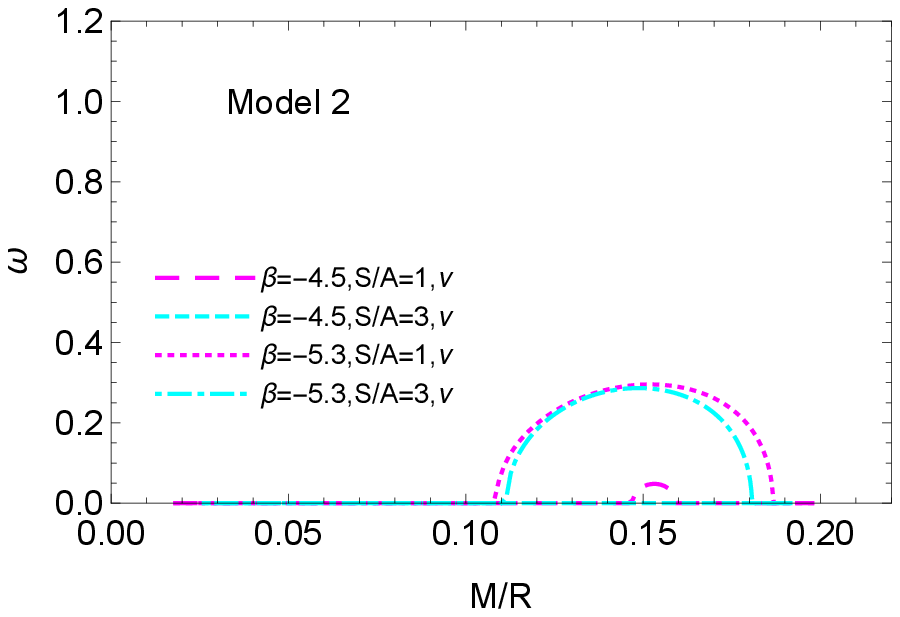}
			\label{phic-rho-2}}
		\caption{Scalar charge versus the compactness $M/R$ for cold NSs and also for PNSs
at finite temperature and with hot isoentropic $\beta$-stable neutron star matter without neutrino trapping and with neutrino trapping for different values of entropy per baryon $S/A$ and the coupling constant, $\beta$, in two models for the coupling function.}
		\label{Scalarcha}
	\end{figure}

In Fig. \ref{Scalarcha}, the scalar charge for cold NSs and PNSs has been presented.
At each value of the compactness, the scalar charge of the PNSs at finite temperature is greater than the one of cold NSs. This difference is more obvious for smaller values of the compactness and also for higher values of $\beta$. Considering the PNSs with hot isoentropic $\beta$-stable neutron star matter, the scalar charge is higher at lower entropies. Besides, the effects of the PNS entropy on the scalar charge are more important when the compactness of PNS is larger.
Our results confirm that neutrino trapping leads to the reduction of the PNS scalar charge similar to the values of $\phi_c$.
With lower values of the coupling constant, $\beta$, the scalar charge becomes larger. This phenomena is more significant in Model 1.

	\section{SUMMARY AND CONCLUDING REMARKS}\label{s7}

The structure of proto-neutron stars in the scalar tensor theory of gravity has been explored applying different proto-neutron star equations of state.
Describing hot pure neutron matter by chiral sigma model and hot $\beta$-stable neutron star matter without neutrino trapping and with neutrino trapping using the finite temperature extension of the Brueckner-Bethe-Goldstone quantum many-body theory, we study the spontaneous scalarization in proto-neutron stars.
Finite temperature as well as the values of the entropy affects the mass of proto-neutron stars.
In the stars at finite temperature, the spontaneous scalarization is more important compared to the stars with hot isoentropic $\beta$-stable neutron
star matter.
Critical central densities at which the spontaneous scalarization starts or terminates can be affected by temperature, entropy, and neutrino trapping in proto-neutron stars.
We have shown that the proto-neutron star equations of state can alter the scalar field and the scalar charge in these stars.

\acknowledgements{The authors wish to thank Dr. Domenico Logoteta for his useful guidance.
We also thank the Shiraz University Research Council.}

\end{document}